\documentclass[preprint]{aastex}
\begin{document}

\newcommand{\up}[1]{\ifmmode^{\rm #1}\else$^{\rm #1}$\fi}
\newcommand{\zdot}{\makebox[0pt][l]{.}}
\newcommand{\upd}{\up{d}}
\newcommand{\uph}{\up{h}}
\newcommand{\upm}{\up{m}}
\newcommand{\ups}{\up{s}}
\newcommand{\arcd}{\ifmmode^{\circ}\else$^{\circ}$\fi}
\newcommand{\arcm}{\ifmmode{'}\else$'$\fi}
\newcommand{\arcs}{\ifmmode{''}\else$''$\fi}

\title{The Araucaria Project. Near-Infrared Photometry of Cepheid Variables
in the Sculptor Galaxy NGC 300
\footnote{Based on  observations obtained with the ESO VLT for Large Programme
171.D-0004
}
}

\author{Wolfgang Gieren}
\affil{Universidad de Concepci{\'o}n, Departamento de Fisica, Astronomy Group,
Casilla 160-C, Concepci{\'o}n, Chile}
\authoremail{wgieren@astro-udec.cl}
\author{Grzegorz Pietrzy{\'n}ski}
\affil{Universidad de Concepci{\'o}n, Departamento de Fisica, Astronomy
Group, Casilla 160-C, Concepci{\'o}n, Chile}
\affil{Warsaw University Observatory, Al. Ujazdowskie 4, 00-478, Warsaw,
Poland}
\authoremail{pietrzyn@hubble.cfm.udec.cl}
\author{Igor Soszy{\'n}ski}
\affil{Universidad de Concepci{\'o}n, Departamento de Fisica, Astronomy Group, 
Casilla 160-C, Concepci{\'o}n, Chile}
\affil{Warsaw University Observatory, Al. Ujazdowskie 4, 00-478, Warsaw,
Poland}
\authoremail{soszynsk@astro-udec.cl}
\author{Fabio Bresolin}
\affil{Institute for Astronomy, University of Hawaii at Manoa, 2680 Woodlawn 
Drive, 
Honolulu HI 96822, USA}
\authoremail{bresolin@ifa.hawaii.edu}
\author{Rolf-Peter Kudritzki}
\affil{Institute for Astronomy, University of Hawaii at Manoa, 2680 Woodlawn 
Drive,
Honolulu HI 96822, USA}
\authoremail{kud@ifa.hawaii.edu}
\author{Dante Minniti}
\affil{Pontificia Universidad Cat{\'o}lica de Chile, Departamento de Astronomia y Astrofisica,
Casilla 306, Santiago 22, Chile}
\authoremail{dante@astro.puc.cl}
\author{Jesper Storm}
\affil{Astrophysikalisches Institut Potsdam, An der Sternwarte 16, D-14482 Potsdam, Germany}
\authoremail{jstorm@aip.de}

\begin{abstract}
We have obtained deep near-infrared images in J and K filters of three fields in the
Sculptor galaxy NGC 300 with the ESO VLT and ISAAC camera. For 16 Cepheid variables
in these fields, we have determined J and K magnitudes at two different epochs, and
have derived their mean magnitudes in these bands. The slopes of the
resulting period-luminosity relations are in very good agreement with the slopes
of these relations measured in the LMC by Persson et al. Fitting the LMC slopes
to our data, we have derived distance moduli in J and K. Using these values
together with the values derived in the optical V and I bands in our previous work,
we have determined an improved total reddening for NGC 300 of E(B-V)=0.096 $\pm$ 0.006 mag, 
which yields
extremely consistent values for the absorption-corrected distance modulus
of the galaxy from VIJK bands. Our distance result for NGC 300 from this
combined optical/near infrared Cepheid study is $(m-M)_{0}$=26.37 $\pm$ 0.04 (random)
$\pm$ 0.03 (systematic) mag and is tied to an adopted true LMC distance modulus of 18.50 mag.
Both random and systematic uncertainties are dominated
by photometric errors, while errors due to reddening, metallicity effects and
crowding are less important. Our distance determination is consistent with the earlier
result from near-infrared (H-band) photometry of two Cepheids in NGC 300 by Madore et al.,
but far more
accurate. Our distance value also agrees with the HST Key Project result of Freedman et al.,
and with the recent distance estimate for NGC 300 from Butler et al. from the TRGB I-band magnitude 
when our improved reddening is used to calculate the absorption corrections. Our distance results
from the different optical and near-infrared bands indicate that the reddening law in NGC 300
must be very similar to the Galactic one.

With the result from this work, the distance of NGC 300 relative to the LMC seems now determined
with an accuracy of $\sim$ $\pm$ 3 percent. The distance to this nearby Sculptor galaxy 
is therefore now known with higher accuracy than that of most (nearer) Local Group galaxies.
\end{abstract}


\section{Introduction}

Cepheid variables are arguably the most important stellar standard candles to calibrate
the range of a few kiloparsecs out to some 30 Megaparsecs on the extragalactic distance
ladder. The HST Key Project on the Extragalactic Distance Scale (Freedman et al. 2001)
has used Cepheid photometry in optical V and I bands in some 25 nearby resolved
spiral and irregular galaxies to improve on the value of the Hubble constant. There
are reasons to believe that Cepheids are even better standard candles when they are
used in the near-infrared regime. Yet, only few galaxies have to date Cepheid
observations in the near-infrared, and usually few variables have been observed
in these galaxies
which have not been sufficient to significantly improve on the distance determinations which have
been made in optical bands. A recent example is the work of the HST Key Project
team which used the NICMOS camera onboard HST to obtain follow-up observations
of a number of Cepheids in several of the Key Project galaxies which had previously
been surveyed for Cepheids in VI passbands (Macri et al. 2001).

In previous papers, we have reported on the discovery of more than a hundred Cepheid variables
in the Sculptor Group spiral NGC 300 (Pietrzy{\'n}ski et al. 2002) from an optical wide-field
imaging survey, and we have derived the distance to this galaxy from the VI light curves
of a long-period subsample of these Cepheids (Gieren et al. 2004). This latter work resulted
in a more accurate determination of the distance of NGC 300 than the previous studies of
Freedman et al. (1992, 2001) which only used the limited number of Cepheids with sparsely
sampled light curves known at the time. NGC 300, at a distance of about 2 Mpc, is a key galaxy in our ongoing
{\it Araucaria Project} in which we seek to improve the calibration of the environmental dependences of several
stellar distance indicators (e.g. Gieren et al. 2001) to the point that the distances
to nearby, resolved galaxies can be measured to 5 percent or better. Apart
from Cepheid variables, our group has also been studying the usefulness of blue supergiants
as a spectroscopic distance indicator in this galaxy, with very encouraging preliminary results
(Bresolin et al. 2002; Kudritzki et al. 2003). Recent deep HST/ACS imaging in several fields
in NGC 300 obtained by us in BVI filters will further allow us to obtain an improved estimate of the distance
of NGC 300 from the tip of the red giant branch magnitude. It is by comparing the distances
of a galaxy coming from such independent methods, and doing this for a sample of galaxies with
widely different environmental properties that we hope to filter out and calibrate one or several 
techniques of distance measurement able to yield the desired high accuracy.

While the Cepheid period-luminosity (PL) relation in optical bandpasses is already a powerful
tool for distance determination, especially when the problem of the appropriate reddening corrections is 
minimized by the application of Wesenheit magnitudes (e.g. Fouqu{\'e} et al. 2003; Storm et al. 2004)), 
one can expect that the use of Cepheids in the near-infrared (NIR) will lead to the most
accurate distance work with these variables. This is for three well-known reasons: First, absorption
corrections in the NIR are small in comparison to their sizes in optical passbands, which is
important because Cepheids as young stars are usually found in dusty regions in their
host galaxies; second,
the intrinsic dispersion of the PL relation due to the finite width of the Cepheid instability
strip decreases with increasing wavelength, and is in K only about half of the intrinsic
dispersion of the PL relation in the V band; and third and very importantly, even random-phase 
Cepheid magnitudes in a NIR
band can already produce a distance determination which can compete in accuracy with the one
coming from optical Cepheid photometry using full, well-sampled light curves. This latter
fact is a consequence of the decreasing amplitudes of Cepheid light curves towards the infrared.
Since Cepheid light curves also become increasingly more symmetric and stable
in their shapes from the optical towards the K band, it is possible to determine accurate corrections to
single-phase NIR observations to derive the mean magnitude of a Cepheid in a NIR band with
very good accuracy. This has been demonstrated, for instance, by Nikolaev et al. (2004), and more recently
 by our group (Soszy{\'n}ski et al. 2005) using a different approach. 
 
 Since in the Araucaria Project 
our interest is to boost the accuracy of the
stellar standard candles we are investigating to their highest possible levels, our prime interest
with Cepheid variables is to use them in the near-infrared, and to calibrate the effect environmental 
parameters, most importantly metallicity, might have on the Cepheid NIR PL relations. Our strategy is therefore
to find a substantial number of Cepheids in each target galaxy from wide-field
optical surveys (where they are most easily discovered), and obtain follow-up NIR photometry at one or two
epochs for a selected subsample of Cepheids which covers a broad range in periods to allow
the construction of accurate NIR PL relations. These can then be compared to the corresponding NIR
PL relations in the Large Magellanic Cloud (LMC) which are now very well established in JHK bands by 
the recent work of Persson et al. (2004). NGC 300 is the first galaxy of our project for which
we report such NIR Cepheid photometry; this work will be followed
by similar studies for the other Local Group and Sculptor Group target galaxies in our project. 
The observations, and the reduction and calibration
of the data will be described in section 2.
In section 3 we will derive the PL relations in the J and K bands from our data and use them,
together with the
information from the optical V and I bands, to estimate an improved reddening for NGC 300, and
 to determine an improved distance to
the galaxy. In section 4 we discuss our results, and in section 5 we will summarize the results 
of this study and draw some conclusions.

\section{Observations, Data Reduction and Calibration}
We have obtained NIR images in J and Ks filters for 3 fields in NGC 300 with the VLT and the ISAAC NIR 
camera/spectrometer of the European Southern Observatory on Cerro Paranal. The coordinates of the
field centers are given in Table 1, and the location of the selected fields in the galaxy is shown in Fig. 1.
The data were obtained in service mode on six nights in the period
July 28-September 05, 2003. Each 2.5 x 2.5 arcmin field 
was observed in JKs on 2 {\it different} nights to cover two different pulsation phases for the
Cepheids in the fields, with the aim to improve the determination of their mean magnitudes.
The pixel scale in our chosen setup was 0.148 arcsec/pix. All images were obtained under
sub-arcsec seeing conditions, and for most images the seeing ranged from 0.3-0.5 arcsec, a crucial advantage 
for the photometry in the crowded fields observed.  
Weather conditions were photometric for at least one of the two observations of a given field.
JHK standard stars on the UKIRT system from the list of Hawarden et al. (2001) 
  were observed on the photometric nights together with the target fields in NGC 300 to allow an
accurate photometric calibration of the data. We observed our target fields in a dithering mode, limiting each
individual integration in J and Ks to 30 s and 15 s, respectively. The total exposure time
for a given target field on a given night was 11.5 min in J and 54 min in K, long enough 
to reach even the
fainter Cepheids in our fields previously discovered in our optical survey
 with a S/N high enough to allow photometry with reasonably small random
errors ($\lesssim$ 0.08 mag). In order to achieve an accurate sky subtraction for our frames, 
we regularly observed random
comparison fields located slightly outside the galaxy where the stellar density was very low,
compared to our target fields, using the AutoJitterOffset template.
 Our fields were chosen as to contain 16 Cepheid variables from the catalog of Pietrzy{\'n}ski et al. (2002),
14 of them with periods
longer than 10 days; the longest-period Cepheid in our sample has a period of 83 days. The use of
such long-period Cepheids makes our distance result independent of problems due to the possible
contamination of the sample with overtone pulsators (see section 4).

Sky subtraction and flatfielding of the frames was performed with the ECLIPSE
package. The PSF model was obtained in an iterative way as described by
Pietrzy{\'n}ski, Gieren and Udalski (2002). The final aperture correction
for a given frame was adopted as a median from aperture corrections obtained
for 10-20 carefully selected stars. The typical rms scatter for the
aperture corrections derived in this manner was about 0.02 mag.

In order to calibrate the photometry onto the standard system,
aperture photometry for the standard stars was performed by choosing an
aperture of 14 pixels. Having relatively few standard star observations
on a given photometric night, typically 5, we adopted mean extinction
coefficients for Paranal as published on the ESO webpage, and the color
coefficients established by Pietrzy{\'n}ski, Gieren and Udalski (2002). This way
we derived zero points in the J and K bands from each standard. The final zero
point in a band was adopted as the mean of the zero points obtained from all standards.
>From the residuals of the standard star observations, the zero point uncertainty
of the photometry is estimated to be $\leq$0.04 mag.

\section{The Cepheid Period-Luminosity Relations in J and K}

In Table 2, we present the journal of calibrated J and K magnitudes of the 16 Cepheids we observed. 
For all Cepheids except cep017, cep032 and cep116 two observations at different phases of their
pulsation cycles were obtained.
The V-band phases of the observations were calculated with the ephemeris given in Gieren et al. (2004),
and are given in Table 2. In the last two columns of Table 2, we show the intensity mean magnitudes of the
Cepheids which we derived by applying a correction to the observed single-phase magnitudes. The corrections
to mean magnitudes were calculated using the actual V-band phases, and the
V and I light curve amplitudes of the Cepheids from our previous optical work together with template light
curves in J and K. A detailed description of the
procedure we use to correct the single-phase J- and K-band observations of Cepheid variables to obtain the
corresponding mean magnitudes has been presented in Soszy{\'n}ski et al. (2005), and the reader is
referred to that paper for details. In short, our correction procedure makes use of the known
amplitudes of a Cepheid in V and I to predict the correction to its mean near-infrared magnitude
from an observation obtained at a known V-band phase. Since for all of our present Cepheids
accurate periods and optical light curves have been established from our previous work, we
were able to apply this model to each of the Cepheids in our sample. Soszy{\'n}ski et al. (2005) have
demonstrated that our adopted procedure reproduces the mean J and K magnitudes from single-phase
J and K observations of a Cepheid with a typical accuracy of 0.02-0.03 mag if the light curves of 
this Cepheid are accurately determined in V and I bands, and its period is well established. This accuracy
is superior to that obtained
from the approach of Nikolaev et al. (2004). It is seen from the data in Table 2 that the
mean magnitudes in J and K derived from the two observations at different phases do indeed agree
very well for most of the Cepheids, with differences between the two independent
determinations of the mean J and K magnitudes for each star which are just a few 
hundredths of a mag, which is in the order
of the uncertainties of the single J and K measurements. For a very few Cepheids in the sample, the
agreement between the two independent determinations of the
mean magnitude is slightly worse; this is likely due to a larger uncertainty on their periods, 
and a minor
quality of their V and I light curves, or to the fact that for some Cepheids we have light curves only
in V, but not in I (see Pietrzynski et al. 2002; Gieren et al. 2004).

In Table 3, we show the final, adopted intensity mean J and K magnitudes for our NGC 300 Cepheid sample, obtained
from a weighted mean of the two individual determinations for each star. Also given are the
respective mean errors of these values which are about 0.03 mag for the brightest Cepheids
in the sample and increase slightly toward the fainter, shorter-period variables. The periods
of the Cepheids were taken from our previous optical studies.
In Fig. 2, we
show the J- and K-band PL relations based on the data in Table 3. In this figure, we have retained the
two individual determinations of the mean J and K measurements for each Cepheid 
(except the 3 stars having observations
at only one epoch), in order to demonstrate the very good agreement between the individual
determinations of the mean magnitudes, which again shows the high efficiency of our correction procedure
mentioned before. In Fig. 3, we plot the adopted, final mean magnitudes of the Cepheids against log P.
These are the final J- and K-band Cepheid PL relations determined from our current data for NGC 300.
In our distance determination from these data, we omit the two shortest-period Cepheids in
the sample which have periods less than 10 days. The faintest Cepheid, cep116, has by far the largest
uncertainty on its J and K measurements and is excluded for this reason. It could also be an overtone
pulsator, given its short period of 5.98 days. The star cep104 (P=7.73 days) is strongly overluminous in the
K band, but not so in the J band. A possible explanation for this behavior is the presence
of a bright, unresolved red companion star. Given the very strong (about 1 mag) deviation of this star from
the mean K-band PL relation, far beyond the observed scatter of this relation,
we prefer to exclude it from our analysis.
We retain for the distance determination all other 14 stars which span
a period range from 13.5 to 83.0 days, yielding an excellent period baseline for the PL fits. 
Weighted least-squares
fits to the mean J and K data of these 14 Cepheids yield slopes of the PL relations of $-3.405$ in J
and $-3.253$ in K, with 1~$\sigma$ uncertainties of 0.047 and 0.044, respectively.
These values have to be compared with the slopes of the PL relations
in J and K determined by Persson et al. (2004) for LMC Cepheids, which are $-3.153$ and $-3.261$,
respectively. While in the K band the agreement of the slope of the PL relation defined by 
the Cepheids in NGC 300
and in the LMC is excellent, the agreement in the J band is less good, but there is still
marginal consistency.
In order to derive the distance of NGC 300 relative to the LMC,
we therefore performed weighted least-squares fits to the data points forcing the slopes to the
corresponding LMC values from Persson et al. This yields the following results:
$$ J = -3.153 \log P + (24.254 \pm 0.048)     \hspace*{1cm}    \sigma=0.180\ \mathrm{mag} $$
$$ K = -3.261 \log P + (23.917 \pm 0.049)     \hspace*{1cm}    \sigma=0.188\ \mathrm{mag} $$
     
According to Hawarden et al. (2001), the J and K magnitudes on the NICMOS system, on which
the Persson et al. (2004) data were obtained, are fainter than the UKIRT magnitudes by a color-independent
and constant 
0.034 $\pm$ 0.004 mag in J, and by 0.015 $\pm$ 0.007 mag in K. On the NICMOS system, our 
PL zero points are therefore 24.288 in J, and 23.932 in K. Using the zero points of the
LMC PL relations as determined by Persson et al. (16.336 in J, and 16.036 in K), we then
obtain a difference of the distance moduli between NGC 300 and the LMC of 7.952 in the J band,
and 7.896 in the K band. Assuming an LMC distance of 18.50 mag, we find the distance modulus
of NGC 300 to be 26.452 in J, and 26.396 in K. To correct these values for interstellar absorption,
we adopted the reddening law of Schlegel et al. (1998) which yields ${\rm A}_{\rm J}$=0.902 E(B-V), and 
${\rm A}_{\rm K}$=0.367 E(B-V). While the foreground Galactic reddening toward NGC 300 is only 0.025 mag
 (Burstein and Heiles 1984),
there is clearly some additional dust absorption produced inside NGC 300, visible on HST
images of the galaxy. In our previous determination of the distance to NGC 300 from optical
VI data, we had found evidence for an additional reddening of 0.05 mag (Gieren et al. 2004) by
comparing the distance moduli derived for the galaxy in V, I and the reddening-independent Wesenheit (W)
band. With the distance moduli determined in the near-infrared J and K bands in this study, we can
improve on this value, by fitting the relationship
$$(m-M)_{0} = (m-M)_{\lambda} - {\rm A}_{\lambda} = (m-M)_{\lambda} - {\rm E}_{\rm B-V} {\rm R}_{\lambda}$$
\noindent   
with the known values of the ratios ${\rm R}_{\lambda}$ of total to selective absorption in VIJK bands (3.24, 1.96,
0.902 and 0.367 from Schlegel et al. 1998), and with the reddened distance moduli $(m-M)_{\lambda}$ we
have determined in VIJK (26.670 and 26.577 in V, I (Gieren et al. 2004), and 26.452 and 26.396
in J and K, this paper). The best-fitting relation yields 
$$ {\rm E}_{\rm B-V} = 0.096 \pm 0.006 $$
$$ (m-M)_{0} = 26.367 \pm 0.011 $$
\noindent
from the slope and intersect of the least-squares fit, which is shown in Fig. 4. Extending
the wavelength coverage of our photometry to the near-infrared, we therefore find a total reddening
to NGC 300 which
is slightly larger than the value of 0.075 we had found from the optical study alone. Our best value for the
reddening-corrected distance to NGC 300 from all bands is 26.367 mag, corresponding to 1.88 Mpc.
Its uncertainty will be discussed in the next section.
 The values in Table 4 show that the
reddening-corrected distance moduli of NGC 300 in the VIJK passbands, calculated with E(B-V)=0.096 as derived above, do 
all agree extremely well with our adopted true distance modulus of 26.367 mag.
It is worth to note that the very small dispersion of the points in Fig. 4 seem to indicate that the
reddening laws in NGC 300 and in the Galaxy must be very similar.

\section{Discussion}
 In this section, we discuss how different sources of random and systematic error may affect our distance
 result for NGC 300. We will not discuss the probably largest single systematic uncertainty, the
 distance to the LMC, which is discussed in a number of recent papers, including Feast (2003), Walker (2003),
 and Benedict et al. (2002). In agreement with our previous work, we have adopted $(m-M)_{0}$ (LMC)=18.50. Should
 future work prove that a different value is more appropriate, our distance result can be easily adapted 
 to this new value.

\subsection{Photometry}
An obvious source of random uncertainty is the photometric noise on our magnitude measurements. The relatively
low scatter of the data points about the mean PL relation in both J and K seen in Fig. 3, which is 
comparable to the scatter of the individual magnitudes of LMC Cepheids around the mean relations
(see Persson et al. 2004; their figure 3), demonstrates that the random errors on our magnitude
measurements are small enough to not significantly increase the observed dispersions of the PL relations
in NGC 300. From the data
and their random errors given in Tables 1 and 2, including the calculation of the mean magnitudes of the Cepheids
from data obtained at only 2 epochs, 
we estimate that the effect of the random photometric errors
on the derived distance modulus does not exceed $\pm$ 0.03 mag. The systematic uncertainty of our photometric
zero points in both passbands is estimated to be $\pm$ 0.04 mag, as discussed in section 3 of this paper.

\subsection{Sample Selection and Filling of the Instability Strip}
One issue of concern is the distribution of the Cepheids in our selected sample over the pulsational
instability strip. In principle, and particularly with a small number of Cepheids it could happen
that all stars tend to lie close to the blue or red edge of the instability strip, rather than having positions
randomly distributed across the strip. This could introduce an effect on the zero point of the derived
PL relation, even if the true slope remains unaffected, which would translate directly into
a corresponding systematic error on the distance modulus derived from these data. From this point
of view, a larger number of Cepheids in our sample would have been desirable to reduce the effect
of this possibly non-homogeneous filling of the instability strip.

In Fig. 5, we have plotted the distribution of the Cepheids in NGC 300 over the instability strip
in the V, B-V color-magnitude diagram. The full width of the instability strip on this diagram
is about 0.5 mag, and it is clearly appreciated that the Cepheids selected for the near-infrared
followup in this paper are distributed more or less randomly over the strip, with no  
 systematic concentration towards the blue, or the
red edge. We therefore conclude that a non-homogeneous filling
of the instability strip is not an important source of error in our present study, and estimate
 that the contribution to
the total random error of our derived distance modulus from non-homogeneous filling of the
instability strip for our selected Cepheid sample does not exceed $\pm$0.03 mag.

Regarding the selection of our final Cepheid sample for the distance solution, we already stated why
the two shortest-period Cepheids were omitted from the analysis. Keeping only stars with periods
longer than 10 days secures a sample with reasonably small random photometric errors, and at the same time
assures that there are no overtone pulsators in the sample which would tend to make our
zero points systematically too bright (and hence the distance of the galaxy too small). In order
to investigate the sensitivity of our derived distance modulus on the chosen cutoff
period, we repeated the fits retaining only the 10 Cepheids with periods longer than 20 days. 
The effect of this change of the cutoff period on the
zero points of the PL fits is less than 0.02 mag in both bands, confirming that the choice of the
cutoff period is not a cause of concern in our analysis. In other words, it means that our distance result
is not affected in any significant way by a Lutz-Kelker bias. 

\subsection{Adopted Fiducial PL Relation}
As we have shown, our data define slopes of the PL relation in J and K which are consistent with the
very accurate slopes derived by Persson et al. (2004) for a sample of nearly 100 LMC Cepheids 
with periods ranging from 3 to 100 days. The period range of the NGC 300 Cepheids selected for our present study
is bracketted by the LMC Cepheids which provide the fiducial PL relations, so there is no reason
of a concern referring to a systematic difference of the mean periods of the program Cepheid sample,
and of the Cepheid sample from which the fiducial relations in J and K were obtained. As a matter of fact, the
only truly accurate Cepheid PL relations in J and K bands available at the present time, in any galaxy, are those
measured in the LMC by Persson et al. (2004). We note, however, that the Galactic near-infrared PL relations established
from the near-infrared surface brightness technique as recently
revised by Gieren et al. (2005) do agree very well with the LMC relations of Persson et al., and have similarly
small dispersions.
The 2MASS data, while existing for a larger sample of LMC
Cepheids, are clearly less accurate because they are based on shallow single-phase observations while
the Persson et al. data are based on full J and K light curves constructed from data of smaller
photometric noise than those of the 2MASS survey. The use of the Persson et al. PL relations to
derive the relative distance of NGC 300 with respect to the LMC is therefore the most reasonable
approach, at the present time.

\subsection{Metallicity Effects on the Distance Modulus}
The effect of metal abundances on the slopes and zero points of Cepheid PL relations in optical and near-infrared
photometric passbands is currently a debated issue. In particular, very little {\it empirical} work
has so far been carried out in the near-infrared. In optical bands, evidence has been mounting that the
{\it slope} of the PL relation is independent of metallicity, particularly with the recent result of
Gieren et al. (2005) that the Galactic Cepheid PL relation derived from the infrared surface brightness
technique seems to have a slope consistent with the LMC and SMC PL relations determined by the OGLE-II
project (Udalski et al. 1999; Udalski 2000). The effect of metallicity on the zero points of 
optical PL relations seems to be less well constrained, at the present time, although recent progress
has been made by the work of Storm et al. (2004), and Sakai et al. (2004).
In near-infrared bands, the only accurate PL relations
existing to date are those measured by Persson et al. (2004) for the LMC. Fortunately, the work on 
the metal abundances
of blue supergiant stars in NGC 300 by Bresolin et al. (2002), and more recently by Urbaneja et al. (2005)
has shown that the average metallicity of young stars in NGC 300 is $\sim$ -0.3 dex,
 which is very close to the average metallicity of
Cepheid variables in the LMC, -0.34 $\pm$ 0.15 dex (Luck et al. 1998). It seems therefore reasonable
to assume that by using the LMC PL relations as the fiducial relations, our distance modulus result for NGC 300
should be practically independent of metallicity effects, even if future work should prove that there
{\it is} a measurable effect of metallicity on the slope and/or zero point of the Cepheid PL relations
in J and K. We remark that the determination of the size of such metallicity effects on infrared
Cepheid PL relations is one of the goals of the Araucaria Project which we will address once we
will have measured the J- and K-band Cepheid PL relations for more target galaxies  of different
metallicities.

\subsection{Reddening}
We believe that with our approach to use the reddened distance moduli obtained in V, I, J and K together
with the reddening law of Schlegel et al. (1998) to constrain the total reddening towards NGC300, including
the intrinsic contribution to the reddening, we have obtained a very accurate result for the total
reddening which has improved our previous estimate (Gieren et al. 2004) which was based on optical photometry 
alone. This claim is supported by the excellent agreement of the values for the absorption-corrected
distance moduli which we obtain for each of the bands (see Table 4). Slight possible changes in the
adopted reddening law would not change our results, and their agreement among the different bands, in any
significant way. If we adopted as the true distance modulus of NGC 300 only the result coming
from the two infrared bands, the change would be less than 0.01 mag, and therefore not significant.
We believe that with our accurate measurement of Cepheid mean magnitudes in both optical and
near-infrared bands we are controlling the effect of reddening on our distance result to a point
that any remaining effect is insignificant.

\subsection{Crowding Effects}
In our previous paper on the distance of NGC 300 from optical photometry (Gieren et al. 2004), the effect
on the Cepheid magnitudes due to unresolved companion stars was extensively discussed, and it was
concluded that the distance determination was not significantly affected by this problem. A confirmation
of this conclusion has very recently come from VI photometry of some 25 Cepheids in NGC 300 obtained
with the {\it Hubble Space Telescope} and ACS camera in six fields in NGC 300, which agrees very well 
with the ground-based
photometry of these stars (Bresolin et al. 2005, in preparation). Since our near-infrared images have
an even better spatial resolution and are deeper than the optical images used in our previous study, and our selected
fields for the near-infrared follow-up are not in the very dense central region of NGC 300, 
we expect that crowding is even less a problem in the determination of the J and K magnitudes of our program
Cepheids than it was in the determination of their V- and I-band magnitudes. 

From the previous discussion, we conclude that the value of our derived true distance modulus of NGC 300
has a total random error of 0.04 mag, and a total systematic uncertainty of about 0.03 mag (which
is by $\sqrt (4-1)$ smaller than the $\sim$0.04 mag uncertainty of each of the individual photometric 
zero points in the four
bands we have used for the calculation of the true distance modulus). This estimate for the systematic
error does, however, not include the contribution coming from the uncertainty on the
adopted LMC distance of 18.5 mag, which is clearly the dominant source of systematic error 
on our derived distance to NGC 300. 
As our final result from VIJK photometry of Cepheid variables in NGC 300, we then find a true distance
modulus of NGC 300 of $(m-M)_{0}$=26.37 $\pm$ 0.04(random) $\pm$ 0.03(systematic) mag (excluding the
systematic uncertainty on the LMC distance, which is difficult to estimate, and scaling our
distance result to an adopted LMC true distance modulus of 18.50). 
Our derived distance value is
more accurate than our previous determination of $(m-M)_{0}$=26.43 mag which we had derived from 
photometry in VI bands alone. 

Butler et al. (2004) have recently found a distance modulus for NGC 300 of 26.56 $\pm$ 0.07(random)$\pm$ 0.13
(systematic) from HST-based photometry
of the I-band magnitude of the tip of the red giant branch. However, in their analysis they only
used a Galactic foreground reddening of 0.013 mag to calculate the absorption correction. 
Using our improved value of E(B-V)=0.096 mag, which takes into account the intrinsic reddening
in NGC 300, the Butler et al. value changes to 26.397, in excellent agreement with the value
derived from the Cepheids in this paper.  
In the study of Bresolin et al. (2005, in preparation), we will derive
a TRGB distance to NGC 300 of significantly reduced uncertainty, as compared to the Butler et al. result,
given that our HST I-band images are going
about 2 mag deeper than those of Butler, and were obtained for six ACS fields. This study will
provide a definitive measurement of the TRGB magnitude in NGC 300 which will then be compared
to our Cepheid-based result.

Our new and definitive distance result for NGC 300 from Cepheid variables is also in excellent
agreement with the earlier result of Madore et al. (1987) who
had found a true distance modulus of 26.35 $\pm$ 0.25 mag from near-infrared (H-band) photometry
of two long-period Cepheids in NGC 300.
While the revised NGC 300 Cepheid
distance obtained by the HST Key Project team (Freedman et al. 2001) from optical photometry
of 26.53 $\pm$ 0.07 mag (also tied to a LMC distance modulus of 18.50, as our present result
in this paper) seems at first glance in marginal disagreement with our result, this is not the case.
Freedman et al. used, as Butler et al. (2004) in their TRGB study, only the Galactic foreground
reddening to correct their NGC 300 distance modulus result for interstellar absorption. Using our
improved reddening value derived in this paper, their result changes to 26.31 $\pm$ 0.07, again
in very good agreement with our result from a combination of optical and near-infrared
photometry of the Cepheids in NGC 300. As a conclusion, all recent determinations of the distance
of NGC 300 from Cepheids and red giants agree with our determination within the combined
1 $\sigma$ uncertainties if consistent reddening corrections are applied. It therefore
seems that the {\it relative} distance of NGC 300 with respect to the LMC is now determined
with an accuracy of about 3 percent.

\section{Conclusions}
We have measured accurate near-infrared magnitudes in the J and K bands for 16 Cepheid variables
in NGC 300 with well-known periods and optical lightcurves. Mean magnitudes were derived from
our two-phase observations by using the correction procedure of Soszy{\'n}ski et al. (2005).
Fits to the observed period-luminosity relations were made adopting the slopes derived from 
the LMC Cepheids
by Persson et al. (2004). By combining the values of the distance moduli derived in the optical
VI and near-infrared JK bands, we have determined E(B-V)=0.096 $\pm$ 0.006 as
an accurate value for the total reddening appropriate for
the NGC 300 Cepheids. Applying this reddening value to correct the observed distance moduli
for absorption, we obtain extremely consistent values for the true distance modulus of
NGC 300 from each photometric band. Our result is $(m-M)_{0}$(NGC 300)=26.37 $\pm$ 0.04 (random)
$\pm$0.03 (systematic) mag, or (1.88 $\pm$ 0.05) Mpc. The random error of this result is dominated
by the photometric noise of our observations and the relatively small sample size, whereas the
systematic uncertainty is dominated by the uncertainty on the photometric zero points. Our
systematic uncertainty estimate does not include the contribution from the distance of the LMC,
which we adopted as 18.50 mag. Effects
due to metallicity, reddening and blending are small and do not significantly affect the
accuracy of our distance measurement. Our distance result is also unaffected by a Lutz-Kelker bias
related to the chosen period cutoff in the PL relation.
The insensitivity of our distance
determination to NGC 300 to metallicity is a consequence of near-identical metallicities
of the young stellar populations in NGC 300 and in the LMC. Applying the fiducial LMC PL
relations in J and K to the Cepheids in more metal-poor or metal-rich systems could cause
a systematic effect on the derived distance if the zero points of the PL relations in J
and/or K are affected by metallicity. Similar work on galaxies with a range of metallicities
should produce improved constraints on this important question. Such studies are 
underway in our Araucaria Project. 

Our distance determination for NGC 300 from combined optical/near-infrared photometry
of Cepheid variables is in very good agreement with the value determined from H-band
photometry of 2 Cepheids by Madore et al. (1987), and with the more recent value of
Freedman et al. (2001) from optical photometry when our improved reddening is applied
to correct for interstellar absorption inside NGC 300. Our result is also in agreement
with the distance determined by Butler et al. (2004) from I-band photometry of the TRGB
in NGC 300, if the additional reddening inside NGC 300 we have determined in this paper
is taken into account. It therefore appears that the distance of NGC 300 relative to the LMC
is now determined with an accuracy of about $\pm$ 3\%, mainly due to the improved
accuracy we have been able to achieve in our present work. 

Given the generally very good agreement between the mean J and K magnitudes of the Cepheids
derived from the individual single-phase observations, it will be a better strategy
in the future to observe each field just once, and double the number of observed fields,
and therefore approximately double the number of Cepheids for the determination of
the PL relation of a galaxy. This will help to reduce the random error due to a
non-homogeneous filling of the instability strip without significantly increasing the
photometric noise on the mean magnitudes.

\acknowledgments
We are grateful to the staff on Paranal who conducted the observations reported
in this paper in service mode, with their usual great expertise.
WG, GP, and DM gratefully acknowledge financial support for this
work from the Chilean Center for Astrophysics FONDAP 15010003. 
Support from the Polish KBN grant No 2P03D02123 and BST grant for 
Warsaw University Observatory is also acknowledged.

\begin{figure}[p] 
\vspace*{18cm}
\includegraphics{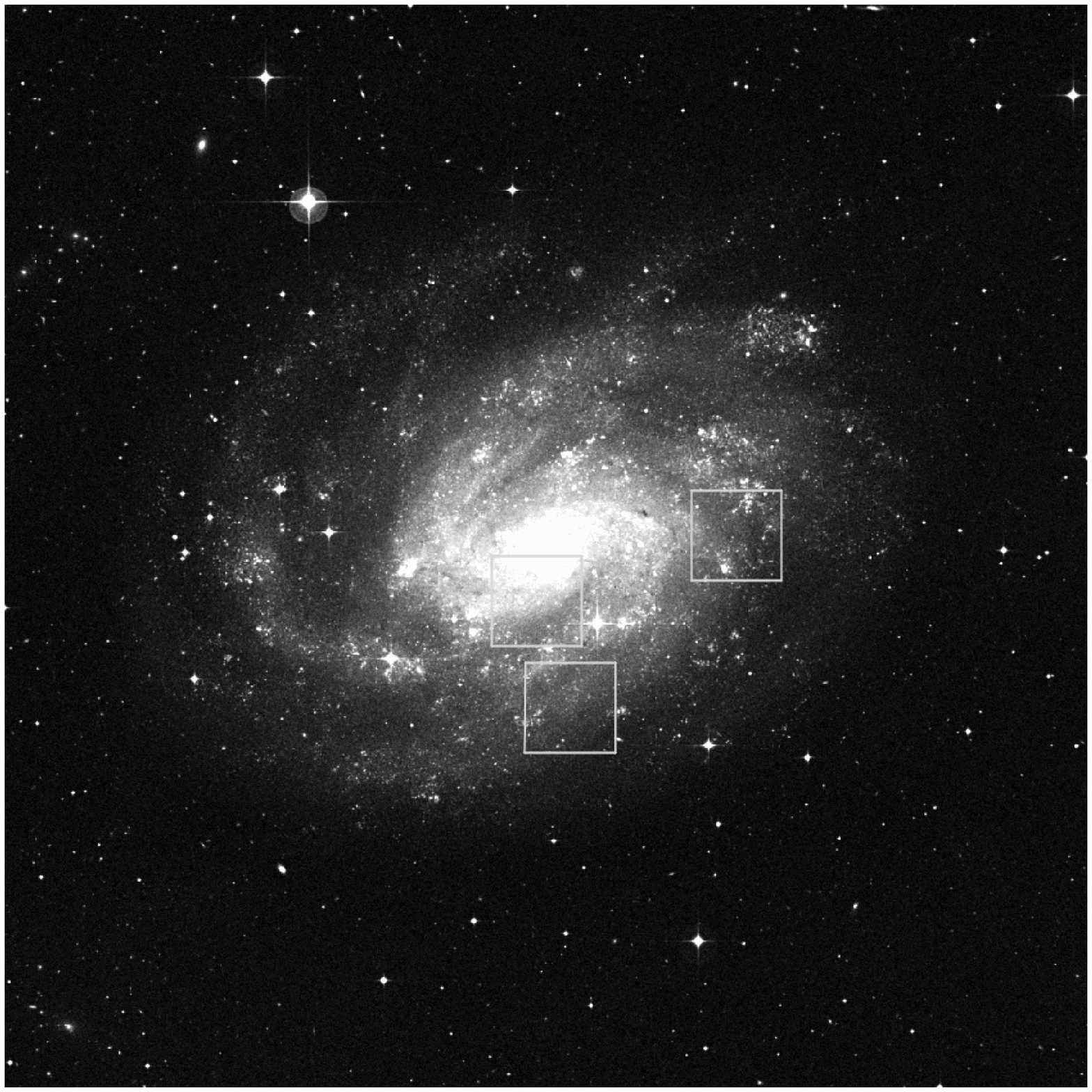} 
\vspace{-3cm}
\caption{The location of the observed VLT/ISAAC fields in NGC300.}
\end{figure}  

\begin{figure}[p]
\vspace*{18cm}
\includegraphics{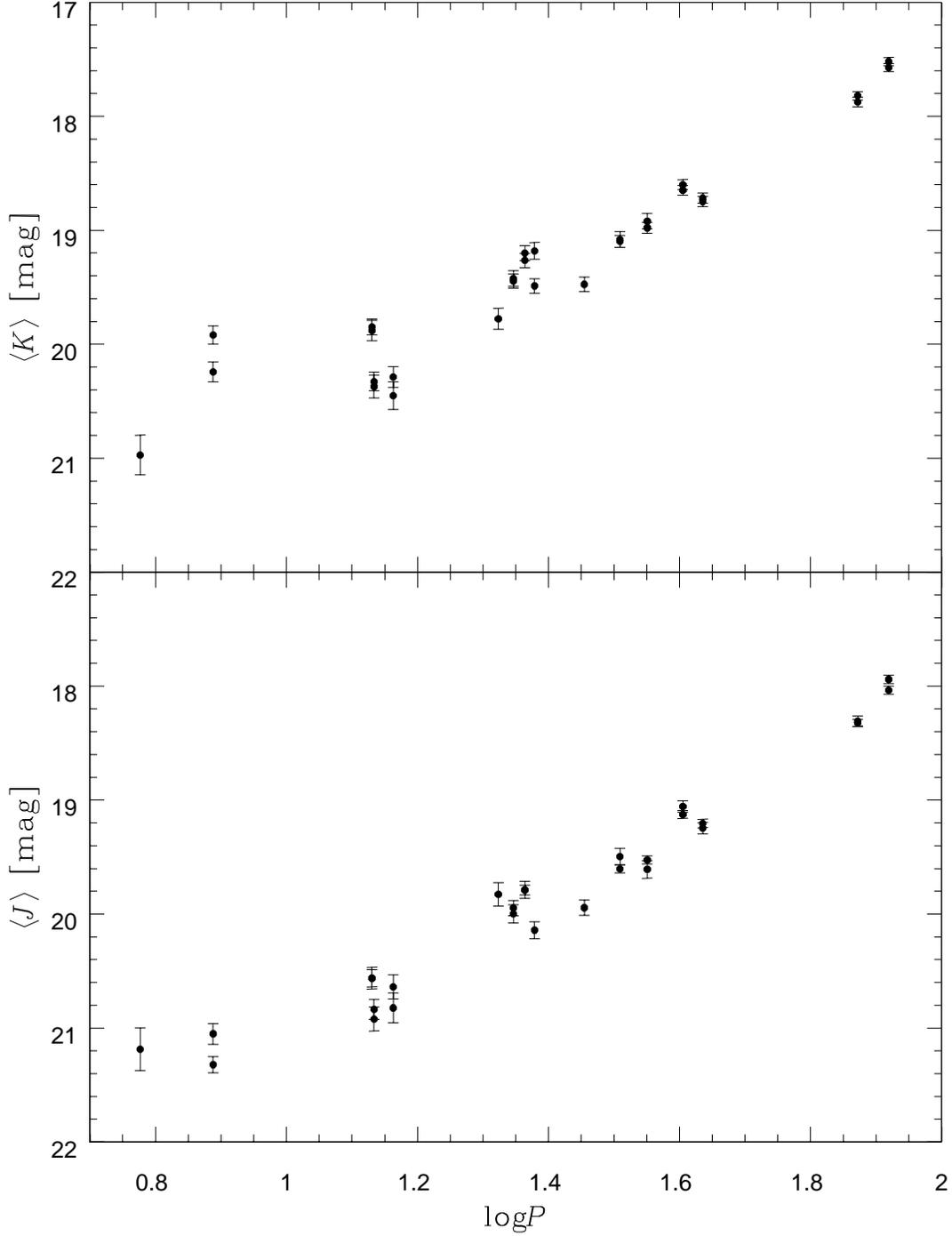}
\caption{The period-luminosity relations in NGC 300 constructed from our data.
         Upper panel: K-band. Lower panel: J-band. For each Cepheid, we show
         the two independent determinations of its mean magnitude from
         observations obtained at different epochs.}
\end{figure}

\begin{figure}[p]
\vspace*{18cm}
\includegraphics{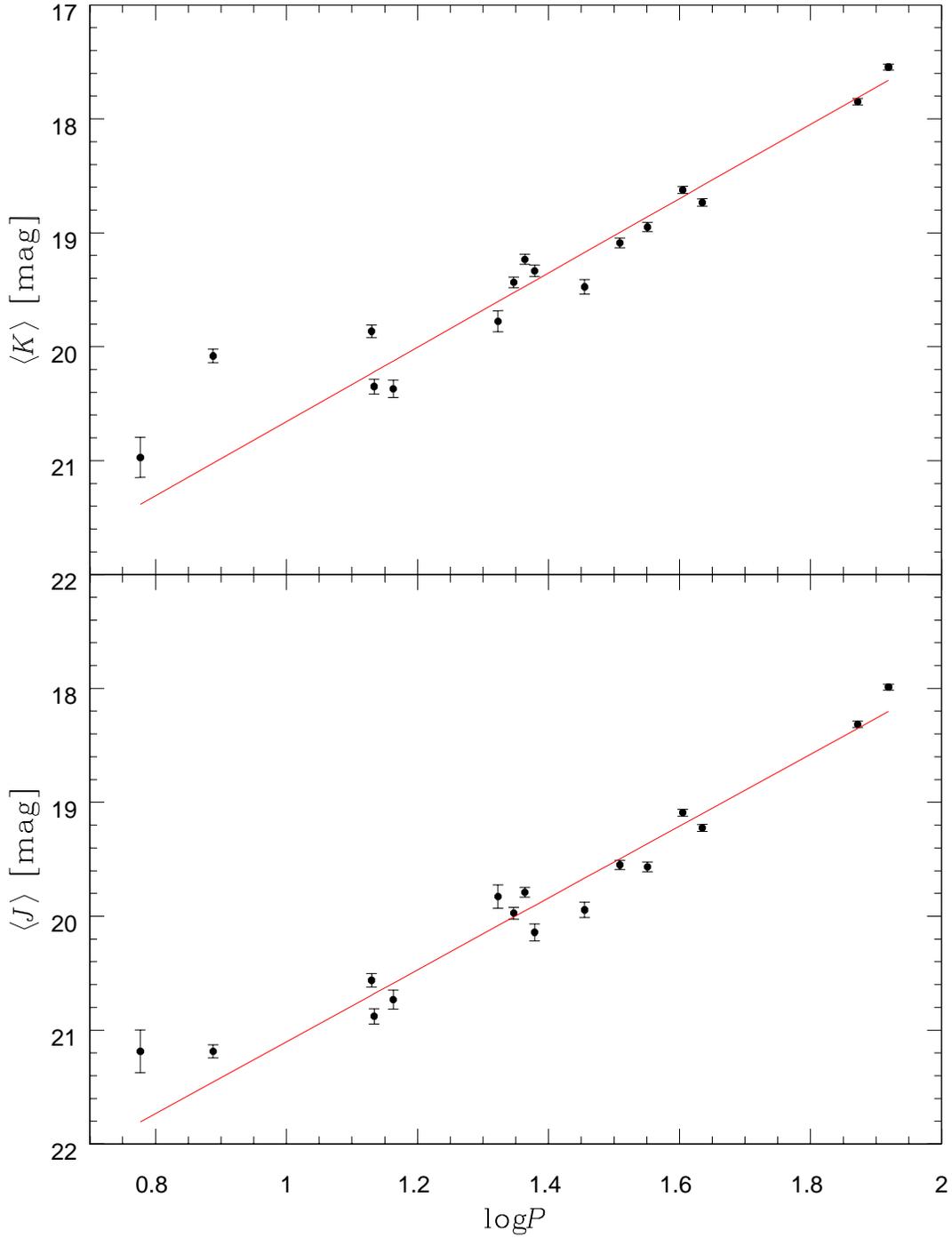}
\caption{The final adopted intensity mean magnitudes of the Cepheids
         in NGC 300, plotted against log P (in days). Overplotted are
         the best-fitting lines for which we have adopted the slopes
         measured by Persson et al. on a large sample of LMC Cepheids.}
\end{figure}

\begin{figure}[p]
\vspace*{18cm}
\includegraphics{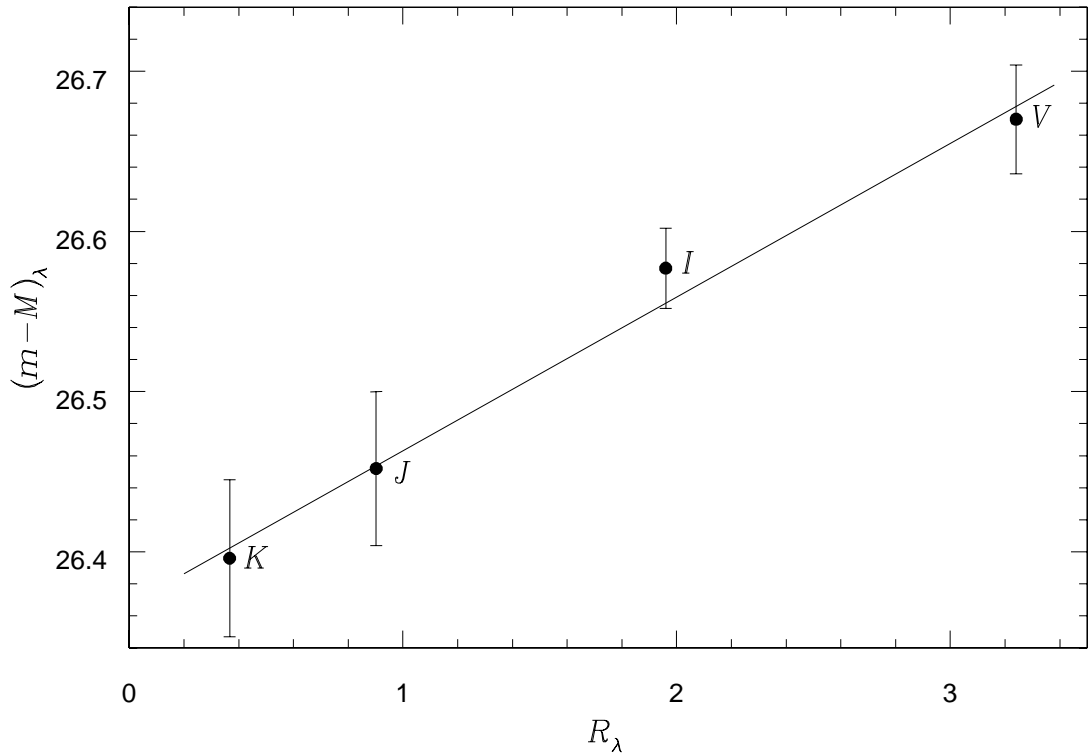}
\vspace{-8.5cm}
\caption{Distance moduli for NGC 300 measured in different photometric
         bands, against the ratio of selective to total absorption for
         these bands. The slope of the best-fitting line gives the
         total color excess, and the intersect gives the true distance
         modulus of the galaxy.}
\end{figure}

\begin{figure}[p]
\vspace*{18cm}
\includegraphics{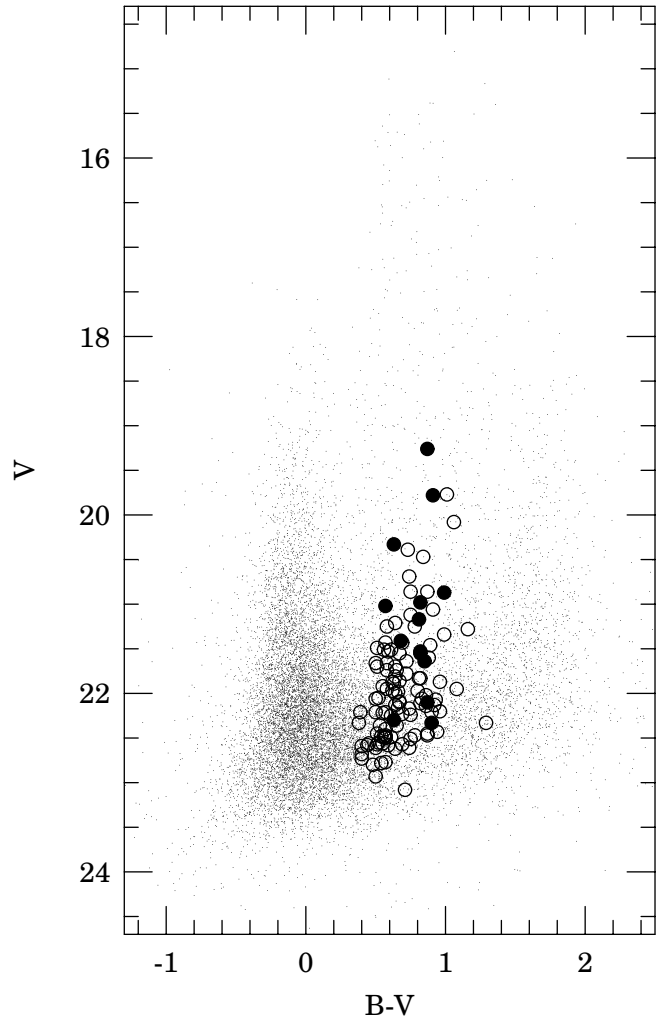}
\vspace{-2.0cm}
\caption{The V, B-V magnitude-color diagram for NGC 300 from our ESO
 photometry (Pietrzy{\'n}ski et al. 2002), with the positions
 of the Cepheid variables overplotted as circles. The filled circles
 indicate the Cepheids with near-infrared J,K mean magnitudes
 in this paper. It is seen that these variables fill the
 Cepheid instability strip near-homogeneously.}
\end{figure}

\clearpage

\begin{deluxetable}{ccc}
\tablewidth{0pc}
\tablecaption{Coordinates (2000.0) of the Centers of the NGC 300 Fields
  observed with VLT/ISAAC}
\tablehead{ & \colhead{RA} & \colhead{DEC}}
\startdata
  Field 1   &  00:54:50.3  &  $-37$:36:31.7 \nl
  Field 2   &  00:54:27.0  &  $-37$:41:19.1 \nl
  Field 3   &  00:54:55.0  &  $-37$:39:30.0 \nl
\enddata
\end{deluxetable}

\clearpage

\begin{deluxetable}{ccccccccccc}
\rotate
\tablewidth{0pc}
\tablecaption{Journal of the Individual J and K Observations of NGC 300
Cepheids}
\tablehead{ \colhead{ID} & $J$ HJD & Phase & $J$ & $\sigma$ & $K$ HJD & Phase & $K$ & $\sigma$ & $\langle{J}\rangle$ & $\langle{K}\rangle$}
\startdata
cep003 & 2452861.88669 & 0.544 & 17.939 & 0.022 & 2452861.79693 & 0.543 & 17.502 & 0.020 & 17.941 & 17.519 \nl
cep003 & 2452873.80867 & 0.688 & 18.069 & 0.021 & 2452873.89290 & 0.689 & 17.591 & 0.018 & 18.036 & 17.572 \nl
cep004 & 2452865.79330 & 0.766 & 18.382 & 0.034 & 2452865.88891 & 0.767 & 17.940 & 0.030 & 18.307 & 17.875 \nl
cep004 & 2452887.77461 & 0.061 & 18.280 & 0.010 & 2452887.68768 & 0.060 & 17.807 & 0.022 & 18.322 & 17.819 \nl
cep007 & 2452848.92857 & 0.938 & 19.306 & 0.022 & 2452848.85179 & 0.936 & 18.868 & 0.034 & 19.205 & 18.718 \nl
cep007 & 2452860.85828 & 0.214 & 19.102 & 0.038 & 2452860.76916 & 0.212 & 18.634 & 0.033 & 19.246 & 18.747 \nl
cep008 & 2452848.92857 & 0.362 & 19.098 & 0.016 & 2452848.85179 & 0.360 & 18.615 & 0.032 & 19.126 & 18.648 \nl
cep008 & 2452860.85828 & 0.658 & 19.094 & 0.040 & 2452860.76916 & 0.656 & 18.624 & 0.033 & 19.055 & 18.600 \nl
cep011 & 2452865.79330 & 0.776 & 19.874 & 0.071 & 2452865.88891 & 0.779 & 19.157 & 0.060 & 19.608 & 18.920 \nl
cep011 & 2452887.77461 & 0.394 & 19.416 & 0.016 & 2452887.68768 & 0.392 & 18.808 & 0.038 & 19.525 & 18.978 \nl
cep015 & 2452865.79330 & 0.918 & 19.679 & 0.067 & 2452865.88891 & 0.921 & 19.309 & 0.061 & 19.495 & 19.081 \nl
cep015 & 2452887.77461 & 0.599 & 19.654 & 0.019 & 2452887.68768 & 0.596 & 19.064 & 0.042 & 19.603 & 19.097 \nl
cep017 & 2452873.80867 & 0.433 & 19.896 & 0.060 & 2452873.89290 & 0.436 & 19.373 & 0.055 & 19.944 & 19.474 \nl
cep025 & 2452848.92857 & 0.822 &   --   &   --  & 2452848.85179 & 0.819 & 19.438 & 0.067 &  9.999 & 19.180 \nl
cep025 & 2452860.85828 & 0.321 & 19.984 & 0.067 & 2452860.76916 & 0.317 & 19.304 & 0.057 & 20.141 & 19.488 \nl
cep028 & 2452865.79330 & 0.586 & 19.819 & 0.069 & 2452865.88891 & 0.590 & 19.243 & 0.055 & 19.788 & 19.264 \nl
cep028 & 2452887.77461 & 0.536 & 19.775 & 0.030 & 2452887.68768 & 0.533 & 19.130 & 0.058 & 19.790 & 19.200 \nl
cep030 & 2452861.88669 & 0.544 & 19.947 & 0.060 & 2452861.79693 & 0.540 & 19.396 & 0.054 & 19.947 & 19.445 \nl
cep030 & 2452873.80867 & 0.081 & 19.912 & 0.074 & 2452873.89290 & 0.085 & 19.399 & 0.063 & 19.998 & 19.424 \nl
cep032 & 2452865.79330 & 0.034 & 19.925 & 0.097 & 2452865.88891 & 0.038 & 19.912 & 0.087 & 19.827 & 19.776 \nl
cep059 & 2452861.88669 & 0.918 & 20.796 & 0.102 & 2452861.79693 & 0.912 & 20.454 & 0.087 & 20.639 & 20.287 \nl
cep059 & 2452873.80867 & 0.737 & 20.970 & 0.126 & 2452873.89290 & 0.743 & 20.544 & 0.118 & 20.823 & 20.450 \nl
cep069 & 2452861.88669 & 0.330 & 20.654 & 0.082 & 2452861.79693 & 0.323 & 20.143 & 0.076 & 20.836 & 20.327 \nl
cep069 & 2452873.80867 & 0.206 & 20.691 & 0.101 & 2452873.89290 & 0.212 & 20.194 & 0.096 & 20.920 & 20.371 \nl
cep072 & 2452861.88669 & 0.325 & 20.460 & 0.069 & 2452861.79693 & 0.318 & 19.744 & 0.062 & 20.563 & 19.847 \nl
cep072 & 2452873.80867 & 0.207 & 20.434 & 0.090 & 2452873.89290 & 0.214 & 19.781 & 0.084 & 20.562 & 19.880 \nl
cep104 & 2452848.92857 & 0.291 & 21.239 & 0.064 & 2452848.85179 & 0.281 & 20.165 & 0.083 & 21.321 & 20.243 \nl
cep104 & 2452860.85828 & 0.835 & 21.176 & 0.086 & 2452860.76916 & 0.824 & 20.028 & 0.073 & 21.050 & 19.918 \nl
cep116 & 2452861.88669 & 0.843 & 21.418 & 0.186 & 2452861.79693 & 0.828 & 21.179 & 0.172 & 21.185 & 20.972 \nl
\enddata
\end{deluxetable}

\clearpage

\begin{deluxetable}{ccccccccccc}
\tablewidth{0pc}
\tablecaption{Final Intensity Mean J and K Magnitudes of NGC 300 Cepheids}
\tablehead{ \colhead{ID} & $\log{P}$ & $\langle{J}\rangle$ & $\sigma$ & $\langle{K}\rangle$ & $\sigma$}
\startdata
cep003 &  1.919 &  17.989 & 0.026 &  17.545 & 0.025 \nl
cep004 &  1.872 &  18.314 & 0.028 &  17.847 & 0.028 \nl
cep007 &  1.635 &  19.225 & 0.031 &  18.733 & 0.032 \nl
cep008 &  1.605 &  19.090 & 0.030 &  18.624 & 0.031 \nl
cep011 &  1.551 &  19.566 & 0.042 &  18.949 & 0.041 \nl
cep015 &  1.509 &  19.549 & 0.041 &  19.089 & 0.043 \nl
cep017 &  1.455 &  19.944 & 0.067 &  19.474 & 0.063 \nl
cep025 &  1.379 &  20.141 & 0.073 &  19.334 & 0.049 \nl
cep028 &  1.364 &  19.789 & 0.043 &  19.232 & 0.045 \nl
cep030 &  1.347 &  19.973 & 0.052 &  19.434 & 0.047 \nl
cep032 &  1.323 &  19.827 & 0.102 &  19.776 & 0.092 \nl
cep059 &  1.163 &  20.731 & 0.084 &  20.368 & 0.076 \nl
cep069 &  1.134 &  20.878 & 0.068 &  20.349 & 0.065 \nl
cep072 &  1.130 &  20.562 & 0.061 &  19.864 & 0.056 \nl
cep104 &  0.888 &  21.186 & 0.058 &  20.081 & 0.059 \nl
cep116 &  0.777 &  21.185 & 0.188 &  20.972 & 0.175 \nl
\enddata
\end{deluxetable}

\clearpage

\begin{deluxetable}{cccccc}
\tablewidth{0pc}
\tablecaption{Reddened and Absorption-Corrected Distance Moduli for NGC 300 in Optical and Near-Infrared Bands}
\tablehead{ \colhead{Band} & $V$ & $I$ & $J$ & $K$ & $E(B-V)$ }
\startdata
 $m-M$                &   26.670 &  26.577 &  26.452 &  26.396 &   --   \nl
 ${\rm R}_{\lambda}$  &   3.24   &  1.96   &  0.902  &  0.367  &   --   \nl
$(m-M)_{0}$           &   26.359 &  26.389 &  26.365 &  26.361 &  0.096 \nl
\enddata
\end{deluxetable}

\end{document}